# A variant of the *h*-index to measure recent performance

Michael Schreiber

*Institute of Physics, Chemnitz University of Technology, 09107 Chemnitz, Germany.*
*Phone: +49 371 531 21910, Fax: +49 371 531 21919*
*E-mail: schreiber@physik.tu-chemnitz.de*

The predictive power of the *h*-index has been shown to depend for a long time on citations to rather old publications. This has raised doubts about its usefulness for predicting future scientific achievements. Here I investigate a variant which considers only the recent publications and is therefore more useful in academic hiring processes and for the allocation of research resources. It is simply defined in analogy to the usual *h*-index, but taking into account only the publications from recent years, and it can easily be determined from the ISI Web of Knowledge.

## 1. Introduction

In 2005 Hirsch defined the *h*-index of a researcher as the largest number $h$ of publications of a scientist which have been cited at least $h$ times. Two years later he claimed that "the *h*-index has the highest ability to predict future scientific achievement" (Hirsch, 2007), because he found a high correlation between the index values of researchers after 12 years and after 24 years of their careers. The observed high product-moment correlation, however, is largely due to a structural correlation which is caused by an order restriction, because the h-index cannot decline and therefore the conventional significance tests are not meaningful (García-Pérez & Núñez-Antón, 2013).

The linear regression fit procedure with elastic net regularization by which Acuna, Allesina, and Kording (2012) were able to predict the time evolution of the *h*-index rather accurately using 18 parameters and still very well with 5 parameters, was immediately discussed by Rousseau and Hu (2012) who raised several objections against the approach. The applicability of the regression equations to other samples was later questioned by Garcia-Perez (2013). Penner, Petersen, Pan, & Fortunato (2013) have also raised doubts about the validity of such predictions in particular for different career-age cohorts and later substantiated these (Penner, Pan, Petersen, Kaski, & Fortunato, 2013) pointing out methodological flaws in the predictions of the *h*-index in general. Comparing the effectiveness of 10 prominent citation indicators Mazloumian (2012) came to the conclusion that the number of annual citations at the time of prediction is the best predictor of future citations.

I have recently demonstrated (Schreiber, 2013a) that the *h*-index is indeed a good predictor of itself, because its increase with time often depends for several years on further citations to rather old publications. Sometimes so-called sleeping beauties which have received few citations for a long time are suddenly discovered and then so frequently cited that they contribute to the *h*-index. This means, however, that its growth does not correlate with the recent performance of a scientist, but it is more likely to result from past achievements (Schreiber, 2013b). Due to this inert behavior it is dangerous to draw conclusions about the future performance from the predictive power of the *h*-index. This raises doubts about its usefulness in hiring processes and project evaluations.

Of course, the described increase of the *h*-index due to long-passée achievements bears testimony of the long-time significance of these old publications. But this does not mean that the scientist is still active and creative and that the *recent* publications still have a large impact (as measured by the number of citations) in the scientific community.

Therefore I study $h_r$ as a variant of the *h*-index which takes into account only the publications from recent years, e.g., the last 6 years or the last 12 years. I presume that the resulting values are more useful to discriminate between currently still successful and influential scientists and not-so-well performing researchers with, respectively, high and low impact of their recent publications. Below I present a case study of the behavior of the Hirsch-type index $h_r$, analyzing the citation distributions of the same 4 examples that I have used previously in my analysis of the predictive power of the *h*-index (Schreiber, 2013a).

A similar proposal has recently been made by Fiala (2014). His so-called h3-index for the year *y* takes into account the publications and citations from the 3 previous years. He discusses the time evolution of that h3-index. Similarly, Pan & Fortunato (2014) have defined a 5-year h-index $h_5$ and used this measure in comparison with their newly proposed author impact factor (AIF). $h_5$ differs from h3, because for its calculation also papers from all earlier years are taken into account, while only the citations are restricted to the recent years as in Fiala's approach and in the analysis below. Fiala (2014) has also proposed other variants labeled h4 and h4' which utilize a 2-year publication window in connection with either a 4-year citation window or a sliding 3-year citation window, respectively. Further, his h3'-index comprises all citation years and therefore corresponds to the index $h_5$ used by Pan & Fortunato (2014). In addition, Pan & Fortunato (2014) have also mentioned a so-called incremental h-index where the calculation is also based only on papers published and citations received in the same time window. But they did not further discuss this suggestion. In particular, neither Fiala (2014) nor Pan & Fortunato (2014) have considered the dependence of their indices on the length of the utilized time window. Therefore the present proposal can be considered a generalization of those suggestions.

## 2. The first example

The citation records for the following investigations were determined in the ISI Web of Science database in September 2014. Care has been taken to establish the integrity of the datasets with respect to homonyms, excluding other authors with the same name as the investigated researchers. In particular, after deselecting authors with different first names as far as these were specified in the database, the paper titles were checked for plausibility, which was not too difficult, because the research fields of the investigated authors are sufficiently close to my own experience. In cases of doubt addresses were checked in the database and compared with the CV where available or alternatively the citation data were compared with the publication list of the investigated author.

The results have been downloaded from the citation reports into a spreadsheet where it is straightforward to sum the citations up to a given year and to count the papers with high citation frequencies up to the value of *h* also selectively, namely depending on the publication year interval as desired. In this way the *h*-index can easily be determined for year *y* considering (only) publications in and after a certain year $r \leq y$ up to and including the year *y*. The year *r* will usually be chosen to be more recent than the beginning of a scientist's career. I shall denote the thus obtained index as $h_r(y)$ in the following. It signifies the $h_r$-index for the year *y*, if the researcher had started publishing in the year *r*. Obviously, for very early years $r_0$ until the beginning of a scientist's career one obtains the usual *h*-index: $h = h_{r_0}(y)$. Values $r > y$ are not meaningful. $h_r(y)$ always increases or remains constant with increasing *y* because more papers and more citations can contribute to the index. On the other hand, it decreases or remains constant with increasing *r* as less and less papers are taken into consideration.

The $h_r$-index is a generalization of the similar indices mentioned in the introduction. In particular, the current index h3 of Fiala (2014) can be written in my notation as h3 = $h_{y-3}(y-1)$. The incremental *h*-index of Pan & Fortunato (2014) is based on the time window $[t-\Delta t,t]$ which in my notation would be given by $r = t - \Delta t$ and $y = t$.

In my previous investigation (Schreiber, 2013a) I had determined the time evolution of the *h*-index for myself and investigated how my *h*-index would have evolved, if I had stopped publishing (or had fallen into sleep) in a certain year *s*, yielding $h_s(y)$. In contrast, now I ask myself the question what the behavior of the *h*-index would be, if I had started publishing (or after sleeping initially, had been roused from sleep) in a certain year *r*. The resulting values of $h_r(y)$ are visualized in Figure 1. Frequently, the lines touch or coincide. But of course, they can never cross.

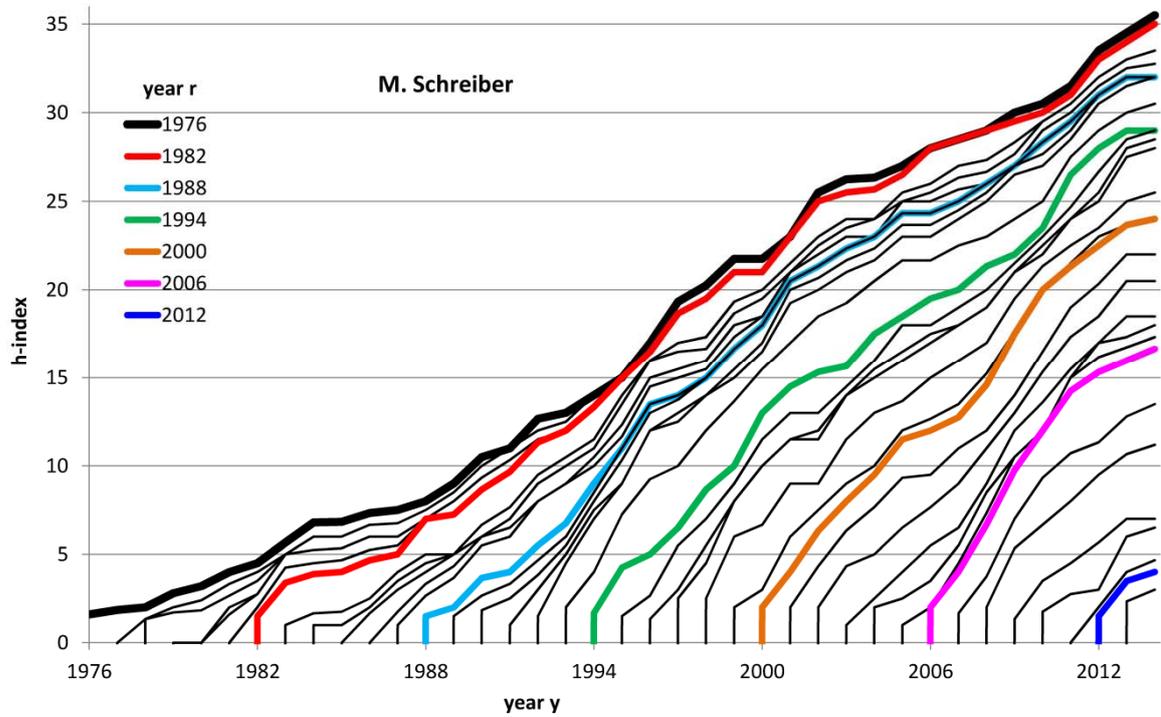

**Figure 1.** Time evolution of the *h*-index for the publications of the present author (top line). Additionally the dependence of $h_r(y)$ is shown starting with the year *r* which is assumed as the year in which the author had started publishing. Selected years (see legend) are highlighted by thick and colored lines. Here the interpolated version of the *h*-index is utilized (Rousseau, 2006, Van Eck & Waltman, 2008, Schreiber, 2008, 2009) that leads to a finer distinction which makes the figure easier to assess.

In Figure 1 I have highlighted certain years *r* in order to make the plot easy to survey. The lines in Figure 1 are more or less dense, i.e. packed or sparse, when considering their crossings with a vertical at some given year *y*. If they are close together this reflects the fact that few publications from these years contribute to my *h*-index, for example for years $1983 < r < 1992$. In principle, it is also possible that lines coincide completely, because a researcher has not published any paper in a certain year. But in my view, this is highly unlikely and in the present datasets it has not occurred. Nevertheless, for very early years $1976 < r < 1982$ the curves in Figure 1 are very close and often even fall onto each other for long time periods. This is not surprising, it only signifies that my early publications did not have a lasting impact. On the other hand one can interpret larger separation between the curves as reflecting years in which I have published papers that have become and remained important for my *h*-index. This is the case for $r = 1982, 1993, 1994, 2006$.

The final points of the curves from Figure 1, i.e. the present values of $h_r(2014)$ are comprised in Figure 2 for all values of *r*. Not all of the above mentioned years *r* in which the separation between the curves in Figure 1 was comparatively large for some years $y < 2014$, are so clearly identifiable as large drops of the curve in Figure 2 as expected. This corresponds to the observation that in several cases the gaps between the curves become somewhat narrower for recent years.

In Figure 2 I have also included the results, which one obtains if only papers were taken into account that belong to the core of $h_{1976}(2014)$, i.e. the 35 papers which at the end of 2014 determined my usual *h*-index. Deleting for this aim those papers from that core which had been published before the year *r* yields the symbols in Figure 2. One can therefore see here how old the publications are which contribute to my present *h*-index. For example, from $r = 1978$ to 1979 one paper drops out of the core and from $r = 1982$ to 1983 three papers drop out of the core. Of course, the thus obtained values are usually smaller than $h_r(2014)$ because for smaller index values other more recent papers become significant and contribute to the core of $h_r(2014)$. In other words, the symbols in Figure 2 indicate how many papers since the year *r* are relevant for my present index value $h = 35$. For example, this means 17 papers since 1997 and 7 papers since 2005. Consequently, the difference between the curve and the symbols in Figure 2 reflects the number of more *recent* papers which contribute to $h_r(2014)$, but have not yet obtained 35 citations and therefore do not qualify for the $h$-core, but only for the $h_r$-core. For example, considering the year $r = 2000$ this means $24 - 10 = 14$ recent papers in the core.

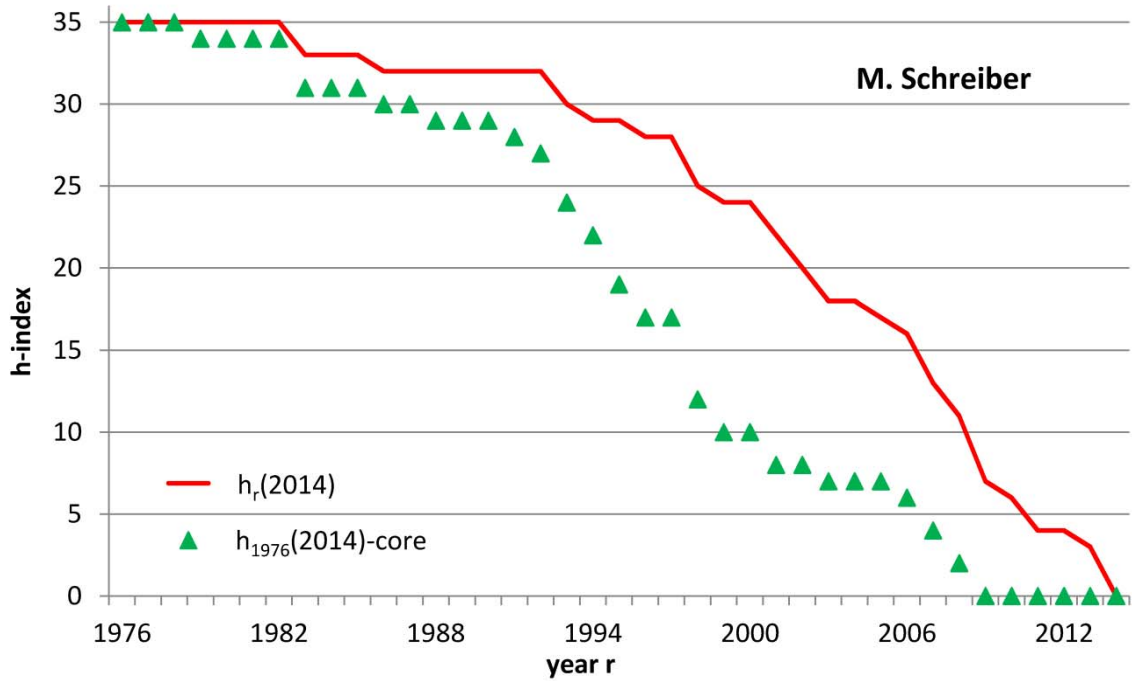

**Figure 2.** Values of the $h_r(2014)$-index for the present author (full line), i.e. in the year 2014 but restricted to the publication activity from the year $r$ onwards. The symbols indicate the $h$-index values if only papers from the $h_{1976}(2014)$-core were taken into consideration, again excluding papers published before the year $r$. As $h_{1976}(2014) = h$ this means the usual core of the $h$-index, i.e. the $h$-defining set of papers. In contrast to Figure 1 the usual non-interpolated (integer) values are utilized, which one can obtain by truncating the interpolated index values.

It is worth noting that in contrast to the evaluation of $h_r(y)$ presented in Figure 1, for the determination of the $h_r(2014)$ data which are shown in Figure 2 it is not necessary to download the citation reports into a spreadsheet and to manipulate them by selecting certain publication years and summing respective citation counts from one year $r$ up to another year $y$ and finally to count the papers with high citation frequencies depending on the desired publication interval. Rather it is much easier, because the information can be directly obtained from the ISI Web of Knowledge by selecting the initial publication years in the search. Then the $h$-index given in the citation report is just the value of $h_r(2014)$.

### 3. Further examples

Like in the previous investigation (Schreiber, 2013a) I have also analyzed the citation records of J. Hirsch, M. Cardona, and E. Witten, see Figures 3, 4, 5, and 6, in order to see whether my own case is typical or whether different behaviors can be distinguished.

Hirsch's values of $h_r(y)$ are displayed in Figure 3. Here the curves for the first 7 years ($r = 1976$ until 1982) are also rather close like for my own data. The next seven lines are less close indicating that a larger number of papers have dropped out of the $h_r$-core. Most prominent is the subsequent large gap between the curves for $r = 1989$ and 1990. This indicates that several papers which are relevant for Hirsch's $h$-index have been published in 1989.

These observations are reflected in the $h_r(2014)$-curve in Figure 4. They are also prominent in the displayed sizes of the $h_{1976}(2014)$-core in that figure. Comparing specifically the $h_r(2014)$ values of Hirsch with my data, one can see after an initial period of constant behavior, a relatively large decrease by 48% within the first 14 years of his career in contrast to a 9% decrease in my case during the same time period. This means that a much larger number of early papers are relevant for Hirsch's $h$-index. This is reflected by the time evolution of the size of the $h_{1976}$-core (see the symbols in Figure 4): until the year $r = 1990$ the $h_{1976}$-core shrinks to 17% in contrast to my case in which it is still 83% in 1990.

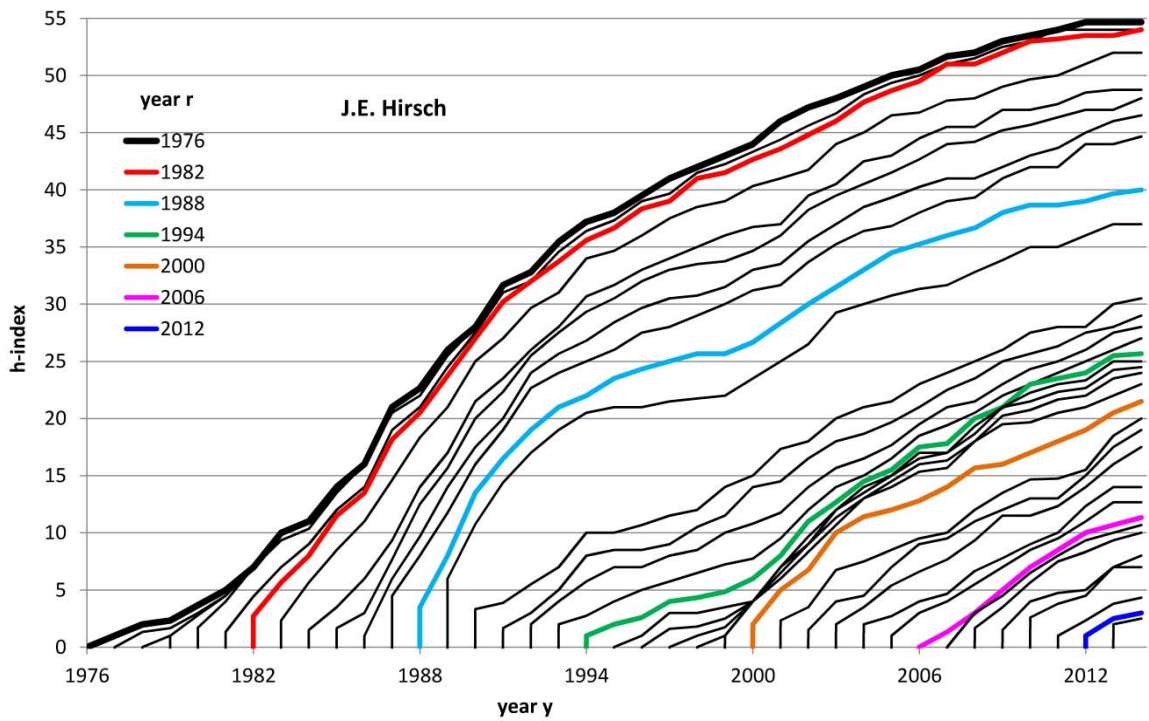

**Figure 3**. Same as Figure 1, but for J. Hirsch.

In consequence of these drastic differences, the subsequent $h_r(2014)$ evolution is nearly the same for Hirsch and myself. This means that the impact of our more recent publications since 1990 is comparably strong. Since 1990 the $h_r$-curves for myself as well as for Hirsch are slightly concave, which means that the decrease becomes stronger in the very recent years.

Looking now at Witten's data one can see in Figure 5 that in contrast to the previous two cases the drop begins immediately in 1976 which signifies papers with very high impact already at the beginning of his career. A strong drop can be observed until 1987, but the absolute values are impressively large. Until 1990 there is a decrease to 57%, which still means $h_{1990}(2014) = 74$. The following decrease is slightly convex and the $h_r$-curve reaches comparable values to the previous two cases in the last 14 years and even lower values in the last 7 years. Consequently the ranking of the investigated researchers changes if only publications from recent years are taken into account.

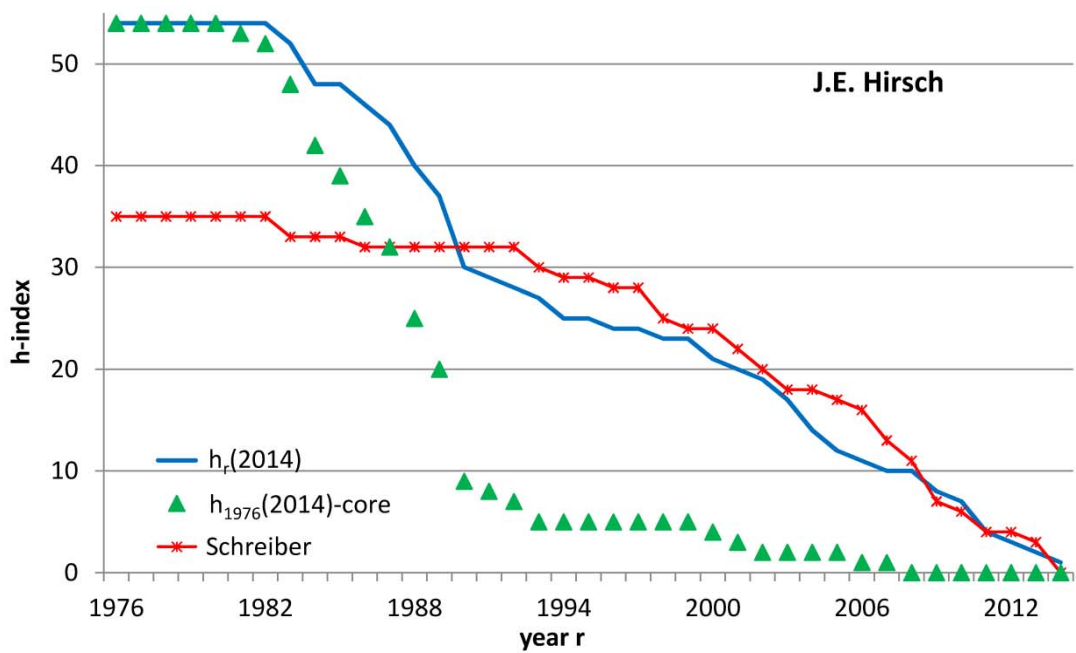

**Figure 4.** Same as Figure 2, but for J. Hirsch. The thin line with stars displays the data from Figure 2 for comparison.

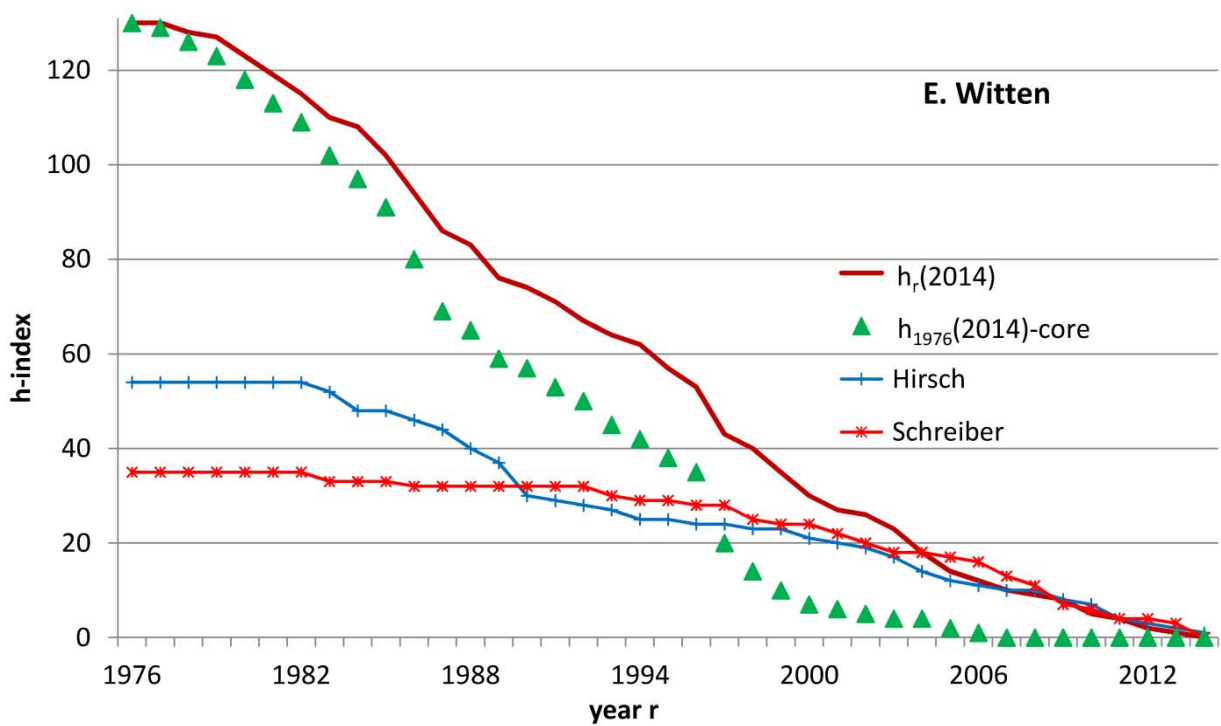

**Figure 5.** Same as Figure 2, but for E. Witten. The thin lines with symbols display the data from Figures 2 and 4 for comparison.

Finally, studying the results for Cardona one can see in Figure 6 a rather slow concave decrease for the entire curve. This means that several early papers are significant for his $h$-index. In the last 12 years it has become comparable to the other three example cases, but of course one should take into consideration that Cardona's career started 18 years earlier. Therefore the high values in the 1990ies are very impressive.

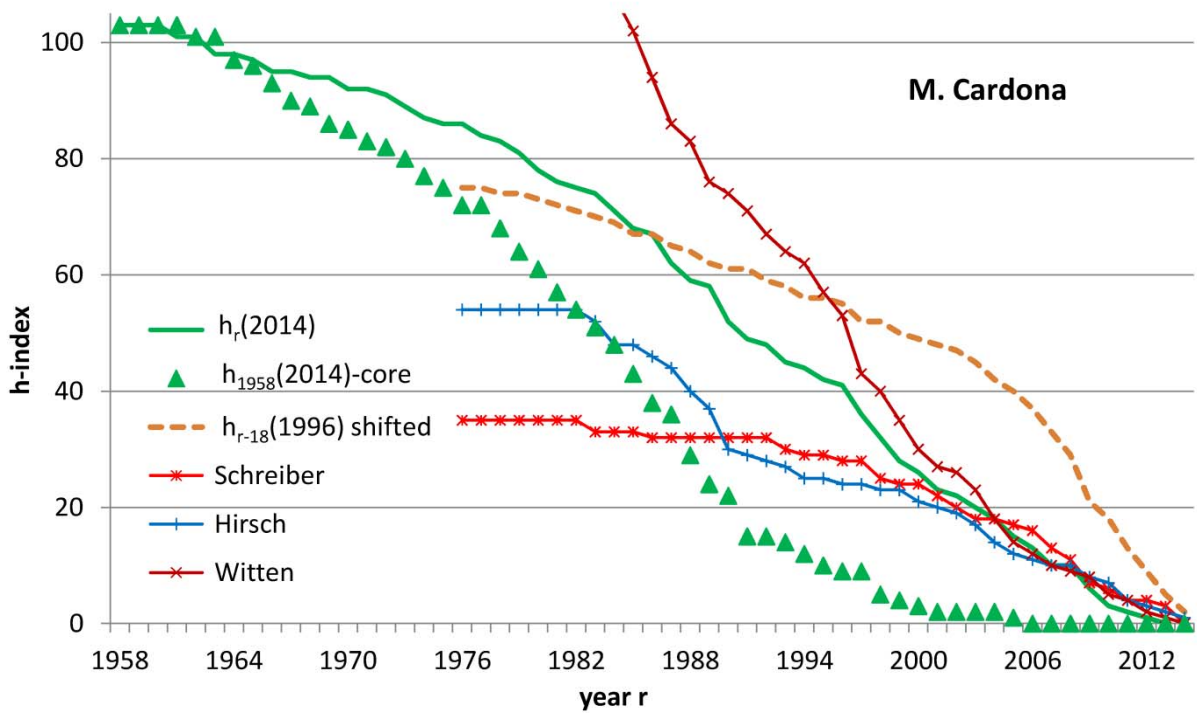

**Figure 6.** Same as Figure 2, but for M. Cardona. The thin lines display the data from Figures 2, 4, and 5 for comparison. The broken line shows the $h_r(1996)$ index, but shifted by 18 years in order to facilitate the comparison which takes into account the different years in which the careers of the 4 scientists started.

In order to compensate for the different career starts, I have also calculated the $h_r(1996)$-values for Cardona. This means that I have taken the citations for the same time spans since the career starts into account as in the other three cases. Shifting the respective curves by 18 years, the data become comparable. The respective broken curve in Figure 6 is different from the other examples, because it shows a rather weak decrease in the first 26 years of his career to still 63% in 1984. As a consequence in the last 14 years the $h_r(2014)$-curves of the previously discussed cases are much lower than (namely about 50% only of) the time-shifted curve for Cardona's values of $h_r(1996)$. The time-shifted curve in Figure 6 thus remains high above my own data and the separation from Hirsch's curve slightly decreases between 1976 and 1982 and then grows steadily until about 2002 and declines after 2005. After 20 years since the starts of their careers Cardona's values become higher than Witten's curve, too. This analysis shows that Cardona has published highly cited papers for a very long time during his career.

Returning the attention to the unshifted curve, I find it impressive that the values are comparable or in most cases even still larger than for the other three scientists in the last 14 years. As mentioned above, considering only publications from the recent years, the ranking is altered. It is not surprising that Cardona's index values are the smallest since $r = 2009$, because he has published rather few papers in recent years.

## 4. Conclusions

Hirsch (2007) has demonstrated the predictive power of the *h*-index by showing that "a researcher with a high h-index after 12 years is highly likely to have a high h-index after 24 years". In my previous investigation (Schreiber, 2013a) I confirmed the claim that the h-index is a good predictor of itself for 4 empirical examples.

As demonstrated now for the same 4 examples the values of the *h*-index result from previous, often rather old publications, i.e., from long passée achievements. Thus these examples demonstrate again that it is dangerous to draw conclusions from the predictive value of the *h*-index with respect to future performance. Therefore, it appears to be dangerous to use the *h*-index in appointment processes or for the purpose of allocating resources.

Of course, any prediction of future performance or impact must be based on passed achievements. Therefore the prediction relies on the assumption that researchers will continue to produce influential papers as they have done in the past. In my view, it is likely that the success of papers in the recent past is more significant than that of papers from the distant past for such a prediction. Therefore, as an alternative I have investigated a variant of the *h*-index, namely $h_r$, which utilizes only the recent publications since the year *r* and the citations to these papers. Thus only the achievements in the last years are evaluated. Consequently the $h_r$-index allows one to distinguish currently still well performing from less successful researchers in terms of the impact of their *recent* work.

For the same aim, Pan & Fortunato (2014) have introduced the AIF, which is defined in analogy to the widely used journal impact factor as the average number of citations given by papers published in year *y* to papers published by the investigated author in the previous years. For the journal impact factor the time period of the last 2 or 5 years is usually chosen. For the AIF the authors applied an interval of 5 years in their investigation. It is not the purpose of the present study to compare the AIF and $h_r$ in detail. Advantages and disadvantages are more or less the same as in comparison of the average number of citations and the standard Hirsch index *h*.

Therefore, one disadvantage of *h*, namely that excess citations (i.e. further citations beyond *h* citations to any of the papers in the *h* core which comprises the *h*-relevant publications of an author) are not taken into account, exists for *h*, as well. This problem has various solutions, for example the multi-dimensional extensions proposed by García-Pérez (2012) or the *g*-index proposed by Egghe (2006) which can be defined as the average number *g* of citations to the *g* most cited papers (Schreiber, 2010). The corresponding definition of $g_r$ for the *g*-index restricted to publications from recent years would also solve another problem that Pan & Fortunato (2014) have mentioned with regard to the incremental *h*-index, namely that the *h*-index is integer and its increments are typically low numbers so that they have little discriminative power. In contrast *g* and $g_r$ are not restricted to integers. On the other hand, one should always be careful to use single number indicators for discrimination between scientists. Small differences of indicator values are not meaningful and therefore it is not reasonable to interpret small differences as reflecting better or worse performance. This applies to $h$ and $h_r$, as well as to the average number of citations and the AIF.

Following the argumentation of the proponents, one advantage of the AIF is that the usually long tail of lowly cited papers keeps the score down. The AIF thus penalizes negligible and/or incremental papers. This is certainly true, but I am not convinced that it is necessary to do so. The authors of those lowly cited papers have already punished themselves, because they have put unnecessary work into writing those manuscripts and they are also in danger of those manuscripts receiving citations which might have been received by their other manuscripts with already higher impact and thus increased the visibility of those papers even more.

From a practical point of view, this consideration of all the lowly cited and uncited papers presents a major problem in the determination of the AIF. It is difficult to determine all the papers of a certain author. Pan & Fortunato (2014) avoided this "tedious problem of disambiguation of authors' names" by analyzing only certain datasets which were already available. It is very laborious to solve this precision problem for the complete set of publications which is needed for the calculation of the AIF. In contrast, for the $h$-index as well as for the $h_r$-index, one only has to check the most cited papers which is usually a much easier task. A more detailed comparison between the AIF and the $h_r$-index and an analysis of the practical differences will be subject of another forthcoming manuscript. A solution of the author disambiguation problem is provided by the ResearcherID. This identification is becoming increasingly popular. For authors with ResearcherID, the data collection in the Web of Science is not hampered by the precision problem and therefore not so laborious, tedious, or difficult.

Further investigations are also necessary to decide what time interval is most reasonable. For established scientists a 12-year time span, i.e., $y - r = 11$ seems to be reasonable. However, for younger scientists this is too long, because it might often extend beyond their career start or include early publications which often have not so much impact. Therefore I suggest to apply an interval of 6 years. In my view 3 years as proposed by Fiala (2014) are too short, because first such a short period may not be representative for an impact of a publication and second the index values are often so small that a distinction between different researchers is not meaningful. A time period of 5 years as employed by Pan & Fortunato (2014) could also be a practically useful time span. In any case, one has to accept that citations take some time to come to the surface due to delays in publication of the citing papers and also due to the schedule with which the database is updated.

With these remarks I do not want to give the impression that I am convinced that the $h$-index or its variants are really meaningful for performance evaluations. My skepticism which I have expressed several times in my previous publications on this topic remains. But it is a fact that nowadays the $h$-index is often used for performance evaluations, in appointment processes as well as for the purpose of allocating resources. With the present investigation I want once more to raise doubts about this practice. Nevertheless, with $h_r$ I have studied a variant which in my view is at least somewhat more useful than the original $h$-index for the evaluation of the recent performance of a scientist. The definition of $h_r$ is as simple as that of the original $h$-index and the $h_r$-index can be easily determined from the citation record.